\documentclass[twocolumn,showpacs,prb,letterpaper,floatfix]{revtex4}%

\usepackage[dvips]{graphicx}%
\usepackage{amsmath}%

\newcommand{\me}{\mathrm{e}}
\newcommand{\mi}{\mathrm{i}}
\newcommand{\md}{\mathrm{d}}
\newcommand{\p}{\emph{\textbf{p}}}
\newcommand{\cvr}{\emph{\textbf{r}}}
\newcommand{\unmedio}{{\textstyle\frac{1}{2}}}

\begin{document}%

\title{Donor centers and absorption spectra in quantum dots}%

\author{Jaime \surname{Zaratiegui Garc{\'\i}a}}%
    \email{jaime.zaratiegui@oulu.fi}%
    \affiliation{Department of Physical Sciences, P.O. Box 3000, FIN-90014 University of Oulu, Finland} %
    \affiliation{Departamento de F{\'\i}sica Te{\'o}rica, Universidad del %
                 Pa{\'\i}s Vasco, Apdo. 644, 48080 Bilbao, Spain}%

\author{Pekka \surname{Pietil{\"a}inen}}%
    \email{pekka.pietilainen@oulu.fi}%
    \affiliation{Department of Physical Sciences, P.O. Box 3000, FIN-90014 University of Oulu, Finland}%

\author{Petteri \surname{Hyv{\"o}nen}}%
    \email{petteri.hyvonen@novogroup.com}%
    \affiliation{Department of Physical Sciences, P.O. Box 3000, FIN-90014 University of Oulu, Finland}%

\date{\today}%

\begin{abstract}
We have studied the electronic properties and optical absorption spectra of
three different cases of donor centers, $D^{0}$, $D^{-}$ and $D^{2-}$, which
are subjected to a perpendicular magnetic field, using the exact
diagonalization method. The energies of the lowest lying states are obtained
as function of the applied magnetic field strength $B$ and the distance $\zeta$
between the positive ion and the confinement $xy$-plane. Our calculations
indicate that the positive ion induces transitions in the ground-state, which
can be observed clearly in the absorption spectra, but as $\zeta\rightarrow 0$
the strength of the applied magnetic field needed for a transition to occur
tends to infinity.
\end{abstract}

\maketitle

\section{INTRODUCTION}

State-of-the-art techniques in the manufacture of semiconductor
devices have made it possible to construct quasi-zero-dimensional
electron systems or quantum dots (QDs), also known as artificial atoms.
These small solid-state systems have similarities with properties
of atoms. Their charge and energy spectra are quantized, and also
they have a full energy level structure~\cite{chakra92}. Because
of this, they are sometimes known as ``artificial
atoms''~\cite{chakra92}, or as ``designer atoms''~\cite{reed93}.
Modern techniques like self-organized growth or molecular beam
epitaxy, allow scientists to fabricate devices as small as 10 nm
and yet their size, shape and other properties can be controlled in
the experiments. Two-dimensional electron systems (2DES) are
confined laterally in between two semiconductor layers. Usually
the materials are GaAs and
$\mathrm{Ga}_{1-x}\mathrm{Al}_x\mathrm{As}$. The comparison
between quantum dots and real atoms is very descriptive, the
nuclear attractive potential is replaced by an artificially
created confinement.

There is an increasing interest in the electronic structure and
properties of donor centers in semiconductors and quantum dots in
magnetic fields. Our work has been focused in studying the cases
of the neutral donor center ($D^0$), negative donor center
($D^-$), which is formed by a neutral donor trapping an extra
electron~\cite{nadja89}, and the case of a neutral donor which
traps 2 electrons ($D^{2-}$). In these quasi-two-dimensional
systems electrons are confined to the $xy$-plane and a positive
ion is located at a distance $\zeta$ from the plane in the $z$
axis. This model was originally proposed by Fox an
Larsen\cite{fox95}. If we restrict to the condition $\zeta=0$ we
are dealing with strictly two-dimensional donor centers. The
greatest interest has been paid into the case of the negative
donor center $D^-$ because it is the one of greater physical
interest. Riva \emph{et al.} considered in a previous theoretical
work~\cite{riva98} the $D^-$ system for quantum well structures.
They found  singlet-triplet transitions as function of the magnetic
field and the parameter $\zeta$.

The negative hydrogenic donor center $D^-$, is the solid-state
analog of the $\mathrm{H}^-$ ion, which consists of a positive ion
with two bounded electrons. This system has attracted considerable
attention since the early days of quantum mechanics. It was as
early as 1929, when Bethe~\cite{bethe29} predicted the stability
of the $\mathrm{H}^-$ ion. Nowadays it is well known that at zero
magnetic field $\mathrm{H}^-$ has only one bound state, while when
magnetic field increases, the ion binds additional states. This
problem has also attracted attention in astrophysics, as the
negative hydrogen ion has been found to be of great importance for
the opacity of the atmosphere of the sun and of similar
stars~\cite{bethe77}.

The existence of $D^-$ centers in center-doped GaAs/Al$_x$Ga$_{1-x}$As 
multiple quantum wells was first reported by Huant, Nadja and 
Etienne~\cite{huant90} in 1990. A system in
which two electrons confined to the $xy$-plane are bound by a
positive ion is called a strictly two-dimensional $D^-$ center.
Fox and Larsen~\cite{fox95} called it $D^-$ barrier center. It is
modeled by generalizing the strictly two-dimensional model by
retaining electron confinement in the $xy$-plane, but moving the
positive ion to a distance $\zeta$ from the plane on the $z$-axis.
In recent years, there has been a lot of theoretical work on the
negative donor barrier~\cite{xie00}. The behavior of the strictly
two-dimensional $D^-$ center in the middle of narrow quantum wells
is known in the strong magnetic field
limit~\cite{larsen92,macdonald92,dzyubenko92}.

In a recent article~\cite{brun02} Brun \emph{et al.} have made near-field
spectroscopy measurements of CdTe/ZnTe single quantum dot at low
temperatures. They found in the photoluminescence spectrum peaks
proving the existence of neutral and negatively charged excitons ($X$ and
$X^-$ respectively) and also multi-excitons. Charged excitons form by hole 
trapping by residual acceptors in the barrier material. They claim that the
signature of the double-negatively charged exciton $X^{2-}$ is present in
the photoluminescence spectra. These latest experiments have motivated us
in considering the calculation of the far infrared (FIR) absorption spectra
for $D^-$ and $D^{2-}$ centers.

$D^-$ centers are one of the simplest many-body systems which
cannot be solved exactly. This system is very interesting in the
study of electron-electron correlations in a two dimensional electron gas
(2DEG), as the screening of impurities in two dimensions in known to have very
drastic effects in the state of the system as, for example, in the quantum Hall
regime.
In our model the separation from the dot to the
positive ion is expressed in units of
$a^{*}_{\mathrm{B}}=\hbar^2\epsilon/m^{*}e^2$ (effective Bohr
radius).

\section{THE EXACT TREATMENT}

We consider a model of donor centers like in earlier works but
generalized for a number of electrons $n_{\me}$. We work with
electron of effective mass $m^{*}$ moving in the $xy$-plane
confined to a parabolic potential and subjected to a perpendicular
magnetic field. In the numerical calculations that follow, we have
used material parameters appropriate for GaAs, \emph{i.e.},
$\epsilon=13$ and $m^{*}=0.067\,m_{\mathrm{e}}$. We have included
spin in our quantum dot calculations but ignored the Zeeman
energy. The confinement potential strength is chosen to be the one of
GaAs dots, \emph{i.e.}, $\hbar\omega_0=4$ meV and the effective Bohr radius is
$a_\mathrm{B}^{\ast}=102.8$ \AA. The Hamiltonian for a donor
center containing $n_{\me}$ electrons is given by
\begin{equation}
{\cal H}={\cal H}_0+{\cal H}_{D}+\frac{1}{2}\frac{e^2}{\epsilon}
\sum_{i\neq j}\frac{1}{|\cvr_i-\cvr_j|},
\label{eqn:Hamiltonian}
\end{equation}
where ${\cal H}_D$ is the donor Hamiltonian given by
\begin{equation}
{\cal H}_D=-\frac{e^2}{\epsilon}\sum_{i=1}^{n_\me}
\frac{1}{\sqrt{r_i^2+\zeta^2}}.
\end{equation}
The term ${\cal H}_0$ stands for the exactly soluble part of the
Hamiltonian, \emph{i.e.}, its eigenfunctions and eigenvalues can be
found analytically. This part is written as
\begin{equation}
{\cal H}_0 = \frac{1}{2m^*} \sum_{i=1}^{n_{\me}}
\left(\p_i-\frac{e}{c} \emph{\textbf{A}}(\emph{\textbf{r}}_i)\right) ^2 +
\frac{1}{2}m^\ast \omega_0^2\sum_{i=1}^{n_{\me}}r^2_i{},
\end{equation}
where the vector potential is in the symmetric gauge,
$\emph{\textbf{A}}(\emph{\textbf{r}})=\frac{1}{2}B(-y,x,0)$,
the single-particle
energies and wave functions can be expressed explicitly as
\begin{eqnarray}
E_{nl}^0 & = & (2n+|l|+1)\hbar\Omega-\frac{1}{2}\hbar\omega_c l,
\label{eqn:energies}\\
\psi_{nl}(r,\theta) & = & \exp(-\mi l\theta) R_{nl}(r),
\end{eqnarray}
where $R_{nl}(r)$ is the radial wave function, given by
\begin{equation}
R_{nl}(r)=C\exp[-r^2/(2a^2)-\mi
l\theta]r^{|l|}L^{|l|}_{n}(r^2/a^2),
\end{equation}
in which $C$ is the normalization constant,
$a~=~\sqrt{\hbar/(m^*\Omega)}$,
$\Omega=\sqrt{\omega_0^2+\omega_c^2/4}$, and $L^{k}_{n}(x)$ is the
associated Laguerre polynomial. It is convenient to define also
the following two parameters, $b=\sqrt{1+4\omega_0^2/\omega_c^2}$
is a dimensionless parameter and $\ell_0=\sqrt{\hbar c/eB}$ is the
magnetic length. The radial and orbital angular momentum quantum
numbers can have the following values
\begin{equation}
n=1,2,\ldots,\ \ l=0,\pm 1,\pm 2,\ldots.
\end{equation}
This single-particle basis is used by the exact diagonalization
method in order to expand the Hamiltonian ${\cal H}$. As ${\cal H}$ is
independent from spin, we can add to our basis set the quantum number
$s_z$. Since the only effect of the spin degree of freedom is to introduce 
exchange terms in the mutual interaction part of the many particle
Hamiltonian~(\ref{eqn:Hamiltonian}), in what follows we omit its explicit
notation.

The exact diagonalization method~\cite{wilkinson88,lin93} consists
in expanding the total Hamiltonian as a series of a given basis and extract
the lowest eigenvalues (energies) of the matrix generated. The
better the basis describes the Hamiltonian, the faster will be the
convergence. A good discussion of the convergence and comparison of this
method was made by Pfannkuche \emph{et al.}~\cite{pfannkuche93}
The most common basis chosen is the one that describes the Hamiltonian
at zero order, \emph{i.e.}, the basis
found from the exactly solvable part of the Hamiltonian, which in
our case is represented by the discrete basis of eigenfunctions $\psi_{nl}$.

When generating the Hamiltonian matrix, the part corresponding to
${\cal H}_0$ will lead to the diagonal
elements~(\ref{eqn:energies}), but the part correspondent to the
donor center is given by the more complicated integral
\begin{eqnarray}
{\cal B}_{n_2l_2}^{n_1l_1}&=&\left<n_1\,l_1|V_D(r)|n_2\,
l_2\right>\nonumber\\
&=&-\frac{e^2}{\epsilon}\int
\psi_{n_1l_1}^*(\cvr)\frac{1}{\sqrt{r^2+\zeta^2}}\psi_{n_2l_2}(\cvr)
\md\cvr.\label{eqn:donor}
\end{eqnarray}
The integration over the angular coordinate $\theta$ gives the
conservation law of the total angular momentum $L=\sum l$.
Therefore, there will be no mixing of energy levels with different
total angular momenta and we will be able to classify the
eigenfunctions by the $L$ quantum number.

If we want to calculate the matrix elements which are different
from zero, we must have $l_1=l_2$. Under these conditions, we can
set $l_1=l=l_2$. We have obtained an exact analytic solution for
the integral which appears in~(\ref{eqn:donor}), ending up with
the analytic solution for the one-particle matrix elements
\begin{eqnarray}
{\cal B}_{n_2l_2}^{n_1l_1} & = & \frac{e^2}{\epsilon}
\frac{b}{\ell_0^2}\left[\frac{n_1!}{(n_1+l)!}
\frac{n_2!}{(n_2+l)!}\right]^{\unmedio} \nonumber \\
 &  & \sum_{\kappa_1=0}^{n_1}\sum_{\kappa_2=0}^{n_2}
 \frac{(-1)^{\kappa_1+\kappa_2}}{\kappa_1!\kappa_2!}
 {n_1+|l| \choose n_1-\kappa_1}{n_2+|l| \choose n_2-\kappa_2}\nonumber \\
 &  & \left(\frac{b}{2\ell_0^2}\right)^{\kappa_1+\kappa_2+|l|}
 \left[{\cal K}_1^n\me^{b\zeta^2/2\ell_0^2}\sqrt{\frac{\pi b}{2\ell_0^2}}
 \mathrm{erfc}\left(\sqrt{\frac{b\zeta^2}{2\ell_0^2}}\right)\right.\nonumber\\
 & & +\left.\sum_{i=2}^n{\cal K}_i^n\zeta^{2i-3}\right],
\end{eqnarray}
where $n=\kappa_1+\kappa_2+|l|+1$, and the coefficients ${\cal
K}_i^n$ are defined recursively in the following form
\begin{eqnarray}
{\cal K}_n^n & = & -\frac{\ell_0^2}{b}, \\
{\cal K}_i^n & = & \frac{\ell_0^2}{b}\sum_{j=i+1}^n
\left[(2j-3){j-2 \choose i-1} \zeta^{2(j-i-1)}\right.\nonumber\\
& & \left.-\frac{2}{r_0^2}{j-1 \choose i-1}\zeta^{2(j-i)}\right]
{\cal K}_j^n.
\end{eqnarray}

We must also calculate the two particle matrix elements,
\emph{i.e.}, the terms corresponding to the electrostatic Coulomb
interaction between the electrons,
\begin{figure}[!t]
\begin{center}
\includegraphics[scale=1]{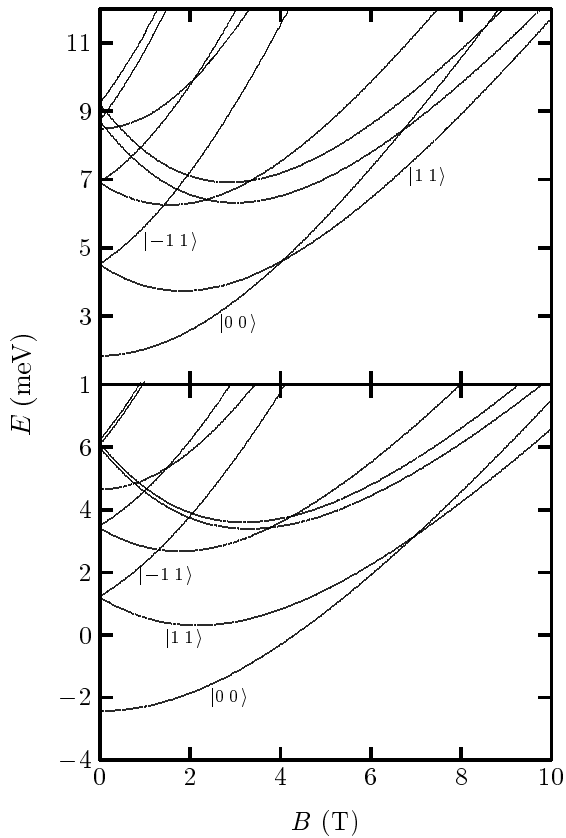}
\end{center}
\caption{\label{E-2e}Lowest energy levels for a donor barrier $D^-$ with two
different values for parameter $\zeta$ as a function of the
magnetic field. Upper picture corresponds to $\zeta=a_{\mathrm{B}}^{\ast}$
and lower picture to $\zeta=\frac{1}{2}a_{\mathrm{B}}^{\ast}$.
Some of the levels are labeled in accordance to two
quantum numbers $|L\,S\left>\right.$.}
\end{figure}
for which an analytic expression
can be found in Ref.~\onlinecite{chakraborty99}.

We have also calculated the intensities of allowed optical
transitions within the
electric-dipole approximation. If we define
the single-particle matrix elements
\begin{eqnarray}
d_{\lambda\lambda'} & = & \left<\lambda'|r\me^{\mi\theta}|\lambda\right> \nonumber \\
 & = & 2\pi\delta_{l+1,l'}\int_0^{\infty}r^2R_{\lambda'}(r)R_{\lambda}(r)\md r,
\label{eqn:dsublambda}
\end{eqnarray}
\begin{figure}[!ht]
\begin{center}
\includegraphics[scale=1]{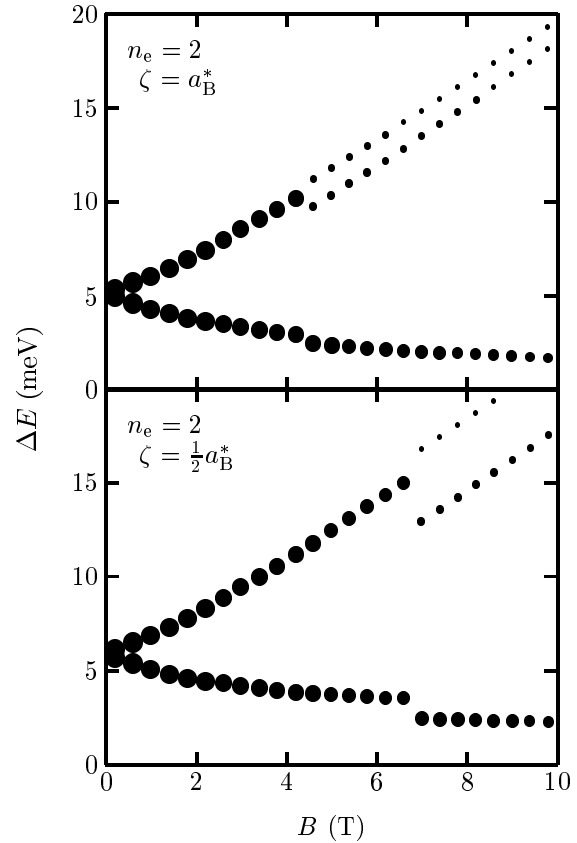}
\end{center}
\caption{\label{FIR-2e}Optical absorption spectra of a donor barrier $D^-$ with
two different values for parameter $\zeta$ as a function of the
magnetic field. The areas of the filled circles are proportional
to the absorption intensities.}
\end{figure}
where $\lambda$ represents the quantum number pair $\left\{
n,l\right\}$ and $R_{\lambda}(r)$ is the radial part of the wave
function corresponding to the state $\lambda$. The dipole
operators can be written as~\cite{halonen96}
\begin{equation}
\left\{\begin{array}{l}
X=\frac{1}{2}\sum_{\lambda\lambda'}\left[d_{\lambda'\lambda}
+ d_{\lambda\lambda'}\right]a_{\lambda'}^\dagger a_\lambda ,\\
 \\
Y=\frac{1}{2\mi}\sum_{\lambda\lambda'}\left[d_{\lambda'\lambda}
-d_{\lambda\lambda'}\right]a_{\lambda'}^\dagger a_\lambda .
\end{array}
\right.
\end{equation}
We will apply this equation for the probability of transition from
the ground state $|0\left>\right.$ to an excited state
$|f\left>\right.$. It can be written as
\begin{equation}
A=|\left<f|\cvr\,|0\right>|^2=|\left<f|X|0\right>|^2+|\left<f|Y|0\right>|^2.
\end{equation}
The areas of the filled circles in Figs.~\ref{FIR-2e} and \ref{FIR-3e}
are proportional to this quantity.

\section{RESULTS and DISCUSSION}

We have made simulations for two different values of the
parameter $\zeta$, $\zeta=a_{\mathrm{B}}^\ast$ and
$\zeta=\frac{1}{2}a_{\mathrm{B}}^\ast$, for both cases, $D^-$ and
$D^{2-}$.

\begin{figure}[!ht]
\begin{center}
\includegraphics[scale=1]{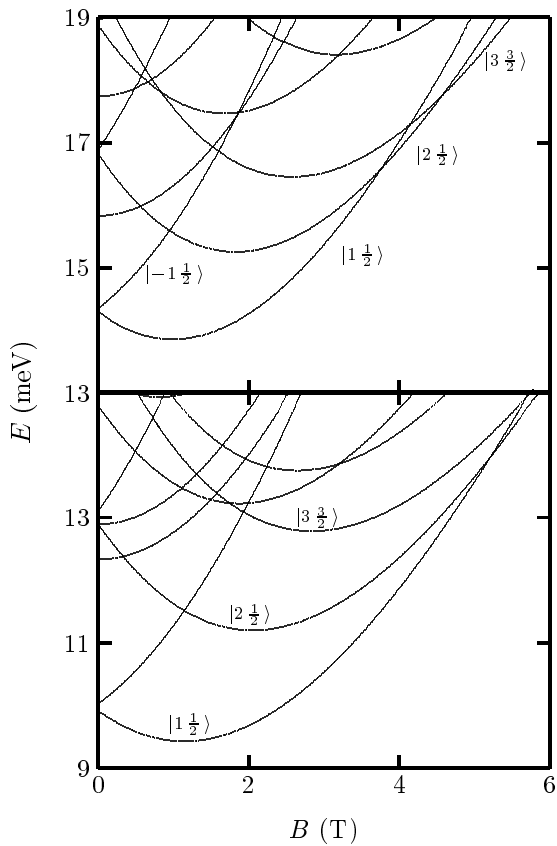}
\caption{\label{E-3e}Lowest energy levels of a donor barrier $D^{2-}$ with two
different values for parameter $\zeta$ as a function of the
magnetic field. Upper picture corresponds to $\zeta=a_{\mathrm{B}}^{\ast}$
and lower picture to $\zeta=\frac{1}{2}a_{\mathrm{B}}^{\ast}$.
Some of them are labeled in accordance to two
quantum numbers $|L\,S\left>\right.$.}
\end{center}
\end{figure}

Figure~\ref{E-2e} shows the lowest lying energy levels for a donor barrier
containing two electrons as a function of the magnetic field. Upper and lower
plots correspond respectively to $\zeta=a^{\ast}_{\mathrm{B}}$ and
$\zeta=\frac{1}{2}a^{\ast}_{\mathrm{B}}$. We can see in both cases
that, at least, a transition in the ground state energy level
close to 4 and 7 Tesla occurs. This fact is also clearly seen in
Figure~\ref{FIR-2e}, in which the dipole-allowed absorption energies are
plotted. There we can observe that at zero magnetic field it is
equally probable for the system to jump to the first and second excited
states, which are degenerate and have $L=\pm 1$. At the point where the
transition occurs, there is a discontinuity in the spectrum, which
is due to the change of the ground state induced by the donor center,
\emph{i.e.}, the spin of the ground state changes from  $S=0$ to $S=1$. Similar
results have been obtained in the FIR absorption spectra of a confined 2DEG
with a Coulomb impurity using Hartree and Hartree-Fock
methods.~\cite{gudmundsson90,gudmundsson94}

\begin{figure}[!ht]
\begin{center}
\includegraphics[scale=1]{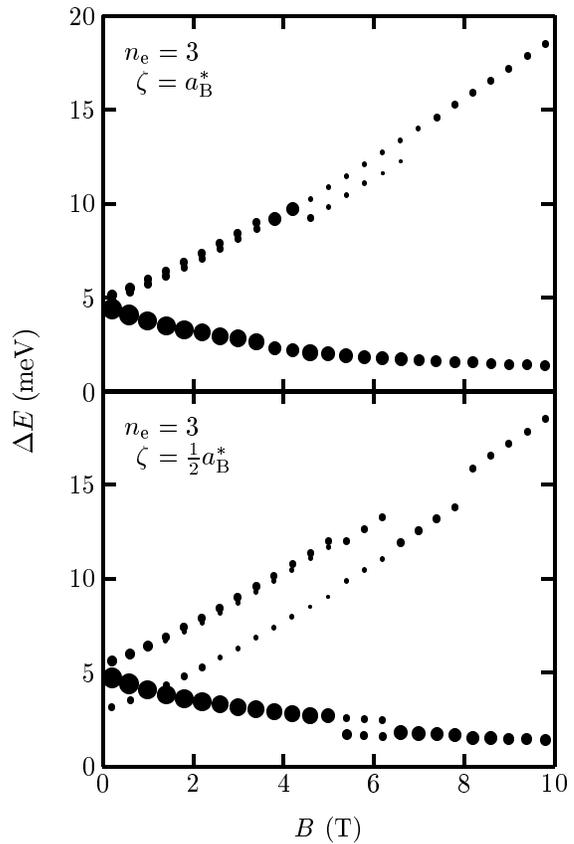}
\caption{\label{FIR-3e}Optical absorption spectra of a donor barrier $D^{2-}$
with two different values for parameter $\zeta$ as a function of
the magnetic field. The areas of the filled circles are
proportional to the absorption intensities.}
\end{center}
\end{figure}

This behavior is also appreciable in the energy spectrum of two and three
interacting electrons without any donor impurity, but unlike the previous
case, the FIR absorption lines do not present any discontinuities nor
splitting. This is due to the fact that two transitions occur simultaneously,
in the ground or initial state and in the excited or final state. The ground
state of the $D^{-}$ and $D^{2-}$ is strongly dependent to the intensity of
the raising magnetic field, increasing its angular momentum. The higher angular
momentum means that the average distance of electrons from the center of the
dot is increased, miminizing the Coulomb interaction between them. The same
effect is also observed in the $X^{-}$ magnetoexciton.~\cite{wojs95}

In Figures~\ref{E-3e} and~\ref{FIR-3e} the energies and the far infrared
transition spectra are shown for the case $D^{2-}$. We observe the same effect
than in the case of the two electrons, \emph{i.e.}, there are induced
transitions in the ground state ($\left.|1\, \frac{1}{2}\right>\rightarrow 
\left.|2\, \frac{1}{2}\right>\rightarrow \left.|3\, \frac{3}{2}\right>$) as
$B$ increases, which is reflected in the FIR
absorption spectra as discontinuities. These spin-singlet--spin-triplet
transitions have been already observed in quantum dots and are known as magic
magnetic number ground state transitions.~\cite{chan01} In this case the
absorption spectra is richer and presents more striking effects, for example,
additional absorption lines appear comparing to the case of three
interacting electrons in a quantum dot. It is convenient to mention that the
optical absorption spectra for the cases of two and three electrons is exactly
the same, this is can be seen as a result of Kohn theorem.

Another effect is the splitting of some of the absorption lines due to some
broken degeneracies in the energy levels. We can also see that the smaller
is the parameter $\zeta$ the larger is magnetic field needed for the transition
to occur. Inspecting the behavior of the transition magnetic field and the
value of $\zeta$, we can expect that for $\zeta=0$ an infinite magnetic field
is needed for a ground state transition to happen.

In our model we used an infinite parabolic potential in order to confine
spatially the electrons, this is not a appropriate selection if we want to
calculate the binding energies of the donor centers $D^{-}$ and $D^{2-}$
because there is no possibility for electrons to jump to an unbound state.
We have set a potential depth according to
references~\cite{ahopelto99,adamowski00,tulkki95}
which allows roughly six bound states in the QD. This gives a binding
energy for the $D^{-}$ of $E_{b}^{D^{-}}\approx 3.8$ meV and for the $D^{2-}$
of $E_{b}^{D^{2-}}\approx -8.2$ meV at zero magnetic field and
$\zeta=a_{\mathrm{B}}^{\ast}$. These results are in accordance with the actual
theories and experimental results for the $D^{-}$ which show the stability
of the charged complex~\cite{larsen92,jiang98}.
The three electron system $D^{2-}$, although it has localized states,
it has a negative binding energy and, therefore, it is unstable toward
the dissociation into a $D^{-}$ and an free electron in the conduction
band.~\cite{dzyubenko93}


\bibliography{articulo}

\end{document}